\documentclass[pra,twocolumn,aps,showpacs,floatfix,superscriptaddress]{revtex4-1}

\usepackage{graphicx}
\usepackage{rotating}
\usepackage{amsmath}
\usepackage{bbm}
\usepackage{subfigure}
\usepackage{color}

\renewcommand{\v}[1]{{\bf #1}}

\newcommand{\be}{\begin{equation}}
\newcommand{\ee}{\end{equation}}
\newcommand{\bea}{\begin{eqnarray}}
\newcommand{\eea}{\end{eqnarray}}
\newcommand{{\br}}{\bf r}

\graphicspath{
  {./}
  {figures/pdf/}
  {figures/eps/}
  {figures/jpg/}
}

\begin{document}


\title{Orbitals from local RDMFT: Are they Kohn-Sham or Natural Orbitals?}

\author{Iris Theophilou}
\affiliation{Peter-Gr\"unberg Institut and Institute for Advanced Simulation,
Forschungszentrum J\"ulich, D-52425 J\"ulich, Germany}
\author{Nektarios N.\ Lathiotakis}
\affiliation{Theoretical and Physical Chemistry Institute, National Hellenic 
Research Foundation, Vass.\  Constantinou 48, GR-11635 Athens, Greece}
\affiliation{Max-Planck-Institut f\"ur Mikrostrukturphysik, Weinberg 2, D-06120 Halle (Saale), Germany}
\author{Nikitas~I.~Gidopoulos}
\affiliation{Department of Physics, Durham University, South Road,  
Durham DH1 3LE, United Kingdom}
\author{Angel Rubio}
\affiliation{Max Planck Institute for the Structure and Dynamics of Matter and Center for Free-Electron Laser Science, Luruper Chaussee 149, 22761 Hamburg, Germany}
\affiliation{Nano-Bio Spectroscopy
group and ETSF Scientific Development Centre, Dpto.\ F\'isica de Materiales,
Universidad del Pa\'is Vasco, CFM CSIC-UPV/EHU-MPC and DIPC, Av.\ Tolosa 72,
E-20018 San Sebasti\'an, Spain}
\author{Nicole Helbig}
\affiliation{Peter-Gr\"unberg Institut and Institute for Advanced Simulation,
Forschungszentrum J\"ulich, D-52425 J\"ulich, Germany}

\begin{abstract}  
\noindent
Recently, an approximate theoretical framework was introduced, called 
local reduced density matrix functional theory (local-RDMFT), where functionals of the one-body reduced density matrix (1-RDM) 
are minimized under the 
additional condition that the optimal orbitals satisfy a single electron Schr\"odinger equation with a local potential. 
In the present work, we focus on the character of these 
optimal orbitals. In particular, we compare orbitals obtained by local-RDMFT with those obtained 
with the full minimization (without the extra condition) by contrasting them against the exact 
NOs and orbitals from a density functional calculation using the local density approximation (LDA). We find that the orbitals from local-RMDFT are very close to LDA orbitals, contrary to those of the full minimization that resemble the exact NOs. Since local RDMFT preserves the good quality of the description of strong static correlation, this finding opens the way to a mixed density/density matrix scheme, where Kohn-Sham orbitals obtain fractional occupations from a minimization of the occupation numbers using 1-RDM functionals. This will allow for a description of strong correlation at a cost only minimally higher than a density functional calculation. 
\end{abstract}

\date{\today}

\maketitle


\section{Introduction\label{sec:intro}}

Reduced-density-matrix-functional theory (RDMFT) \cite{G1975} is an alternative formulation of the many-electron problem where every ground-state property, including the ground-state energy, is a functional of the one-body reduced density  matrix (1-RDM). A main advantage compared to density functional theory (DFT) is that the electronic kinetic energy can be written explicitly in terms of the 1-RDM. Like in DFT, approximations in RDMFT can be cast in a form where all terms are simple explicit functionals of the 1-RDM except a remaining unknown part of the electron-electron interaction term which can be also called exchange and correlation energy. Contrary to DFT the exchange-correlation (xc) energy does not contain kinetic-energy contributions since that part of the energy is treated exactly. Constructing functionals amounts to introducing approximate forms for the xc energy term in terms of the 1-RDM. Most approximations in RDMFT are explicit functionals of -- and are minimized in terms of -- the natural orbitals (NOs), $\phi_j(\v r)$, and their occupation numbers, $n_j$. Various different approximations for the total energy functional have become available over the past decades \cite{M1984,GU1998,bb0,GPB2005,power_finite,pernal2010,AC3,pade,pnof1,pnof2,LHZG2009,PNOF5,sharma08,piris_jcp2013,KP2014,piris_jcp2014} which have proven to describe correctly such diverse properties as molecular dissociation \cite{bb0,GPB2005,power_finite,pernal2010,AC3} or band gaps\cite{sharma08,helbig09,dfg10,SDSG2013}. A major drawback of RDMFT, however, is the increase in computational cost compared to a DFT calculation which is mostly due to the optimization of the NOs. 
Contrary to DFT, the functional variation with respect to the orbitals does not reduce to an iterative eigenvalue problem.
A few techniques have been introduced to define effective Hamiltonian schemes with non-local 
potentials to obtain the natural orbitals\cite{pernal_epot,piris_jcc,Baldsiefen2013114}. Although some of these techniques reduce the computational cost substantially compared  with the full minimization, orbital determination in RDMFT still remains a bottleneck when 
compared with DFT or even Hartree Fock methods.

In an attempt to incorporate the merits of RDMFT functionals, like static correlation effects, in Kohn-Sham-like 
equations an alternative approach, called {\it local-RDMFT}, was introduced recently \cite{localrdmft}.
 The main idea in this approach is to optimize the orbitals under the additional constraint that they are the eigenfunctions of a Hamiltonian containing only a kinetic term and a local scalar potential.
Due to this strong constraint, local-RDMFT does not solve the full 
problem of orbital optimization in RDMFT. Instead, the orbital optimization can be considered as being in the realm of 
DFT, and the orbitals satisfy single particle equations.
 The cost for one optimization of the orbitals in local-RDMFT is similar to the cost of a density functional calculation using an orbital functional via the optimized effective potential method. Local-RDMFT can potentially describe static correlations, where most DFT approaches fail. More specifically, it was found 
that a qualitatively correct dissociation of H$_2$ and N$_2$ molecules is obtained. In particular, for
H$_2$, if the corresponding functional of the 1RDM gives the correct physical picture of dissociation in standard 
RDMFT, then this desirable feature is also preserved by local-RDMFT.

In addition, the calculated energy eigenvalues of the Hamiltonian in local-RDMFT were shown to provide a useful energy spectrum
for molecular systems. It was found that those energy eigenvalues are very good approximations for the ionization potentials (IPs) of a set of atomic and molecular systems \cite{localrdmft,localrdmftappl} 
although a mathematical proof of their association with the IPs is lacking.
In the present work, we refer to RDMFT with the full minimization of the NOs as full-RDMFT and use the prefix full also for specific functionals, e.g.\ full-M\"uller, in order to distinguish from local-RDMFT for which the prefix local will be used. 

\begin{figure}
\centerline{\includegraphics[width=0.40\textwidth, clip]{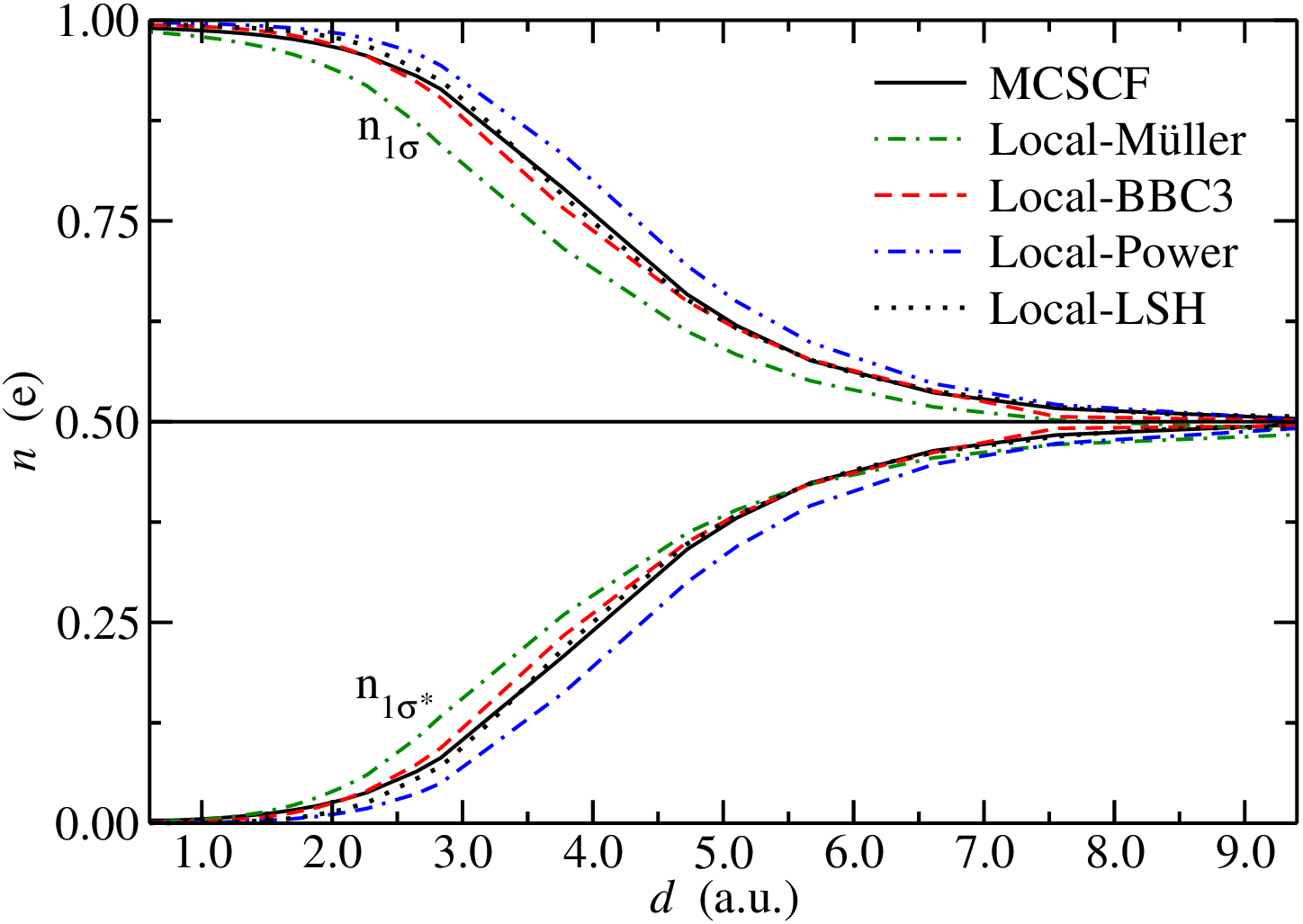}}
\caption{\label{fig:nd}  The dependence of the occupation numbers of the $1\sigma$ and $1\sigma^*$ spin-orbitals
of H$_2$ as a function of the internuclear distance. Only functionals which yield a qualitatively correct dissociation are included (see text for details).}
\end{figure}

A given set of NOs cannot be obtained as the eigenfunctions of a Hamiltonian with a local potential, as we know from their asymptotics. Hence, enforcing the additional constraint is an additional approximation and the optimized local-RDMFT orbitals  (LROs) are expected to be different from the natural orbitals. As a result, the question arises how much the  LROs differ from the orbitals obtained from a full minimization, i.e.\ a minimization of an approximate energy functional without the additional constraint, and from the exact NOs obtained with a very accurate multiconfigurational method. Since the constrained minimization scheme strongly resembles a density functional calculation one might expect the resulting LROs to be closer to the KS orbitals of a DFT calculation. This is certainly true for the asymptotic behavior of the orbitals in finite systems. The exact NOs decay exponentially and if their occupation number is fractional, which is generally the case, they will all share the same exponent for the decay, determined by the chemical potential of the system \cite{MPL1975}. The KS orbitals for a finite system also decay exponentially but the exponent is determined by their KS eigenvalues which implies that the orbitals decay with different exponents unless they are energetically degenerate. The same holds true for the LROs, i.e.\ they decay with an exponent that is determined by the associated energy eigenvalue. This has little effect on the total energy since the asymptotic region contains hardly any density and, hence, does not contribute significantly to the total energy.

In the present work, we focus on the nature of LROs. More specifically, we compare the LROs from an approximate RDMFT functional in the local framework with (a) the orbitals from a full minimization using the same approximate functional, (b) very accurate approximations of the exact NOs obtained with multi-configuration self-consistent field (MCSCF) calculations, and (c) the KS orbitals from a DFT calculation using the local density approximation (LDA). This allows us to investigate if the orbitals from local-RDMFT are indeed closer to KS orbitals than to the exact NOs. 
We shall find that the LROs resemble the KS orbitals. Hence, our work provides the motivation 
for further exploration of the possibility of hybrid DFT/RDMFT approaches, for example to obtain the orbitals from a flavor of DFT and the occupation numbers through a minimization of an appropriate RDMFT functional.

The paper is structured as follows: in section \ref{sec:localrdmft} we review the basic ideas of local-RDMFT and the Schr\"odinger equation that the  LROs have to satisfy. In section \ref{sec:results} we then compare the shapes of the NOs from a full minimization with those from local-RDMFT, using different common approximations for the exchange-correlation energy in RDMFT \cite{M1984, GPB2005, pade, power_finite,LSH1956}, for the helium atom and for the hydrogen molecule at different internuclear distances. To complete the comparison we also include the almost exact NOs from a MCSCF calculation, the KS orbitals from a DFT calculation using the LDA, and the Hartree-Fock (HF) orbitals. We conclude the paper in section \ref{sec:conclusion}.

\begin{figure}
\includegraphics[width=0.4\textwidth, clip]{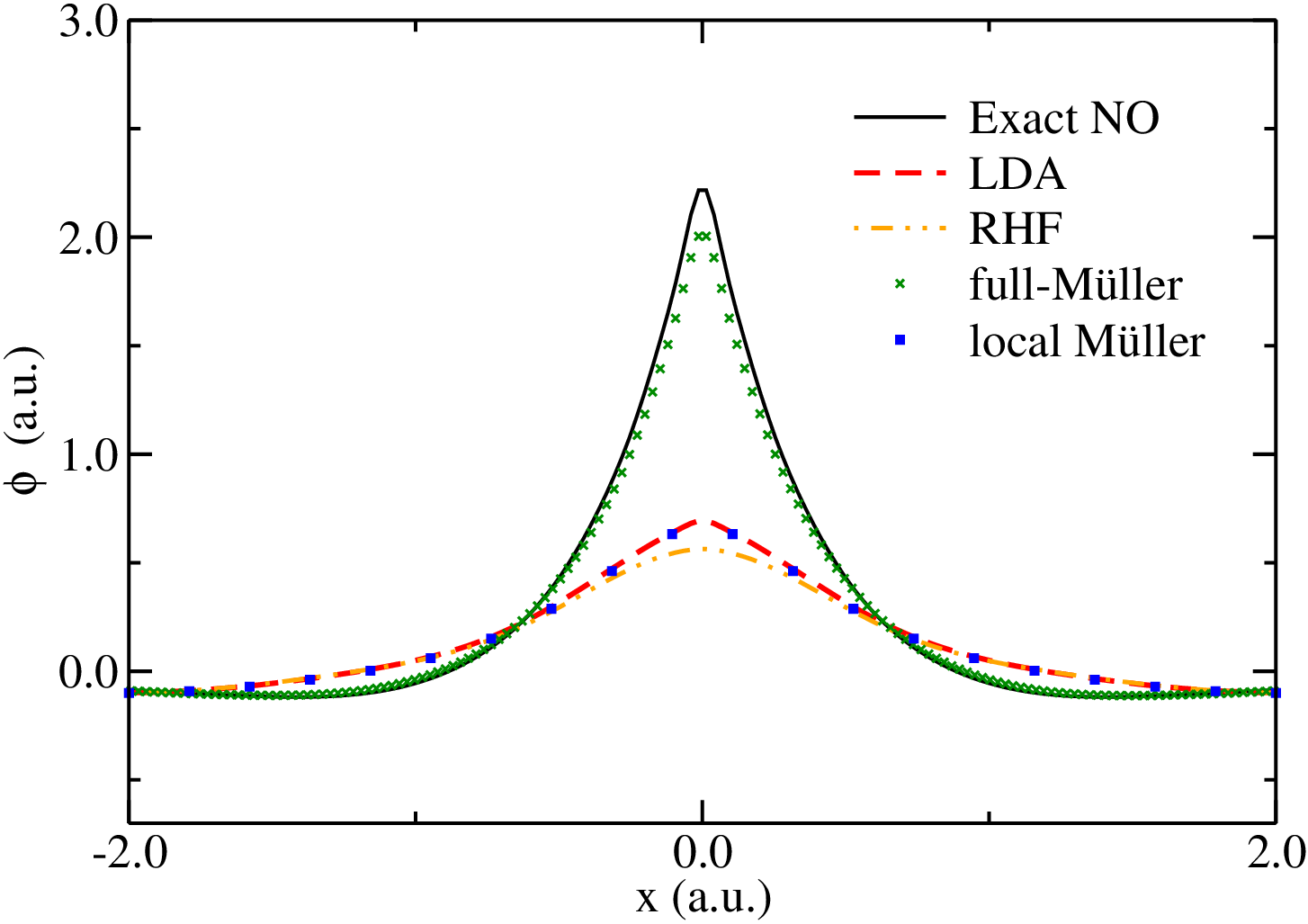}
\caption{\label{fig:He_2nd} Helium 2$s$ orbital along an axis through the position of the nuclei obtained with  
full- and local-RDMFT compared with the exact NO, and those obtained with LDA and RHF.
The orbital obtained with either local-BBC3, local-Power or local-ML is very similar to the one given by local-M\"uller.} 
\end{figure}

\section{Local RDMFT}\label{sec:localrdmft}

\begin{figure*}
\begin{tabular}{cc}
\includegraphics[width=0.45\textwidth, clip]{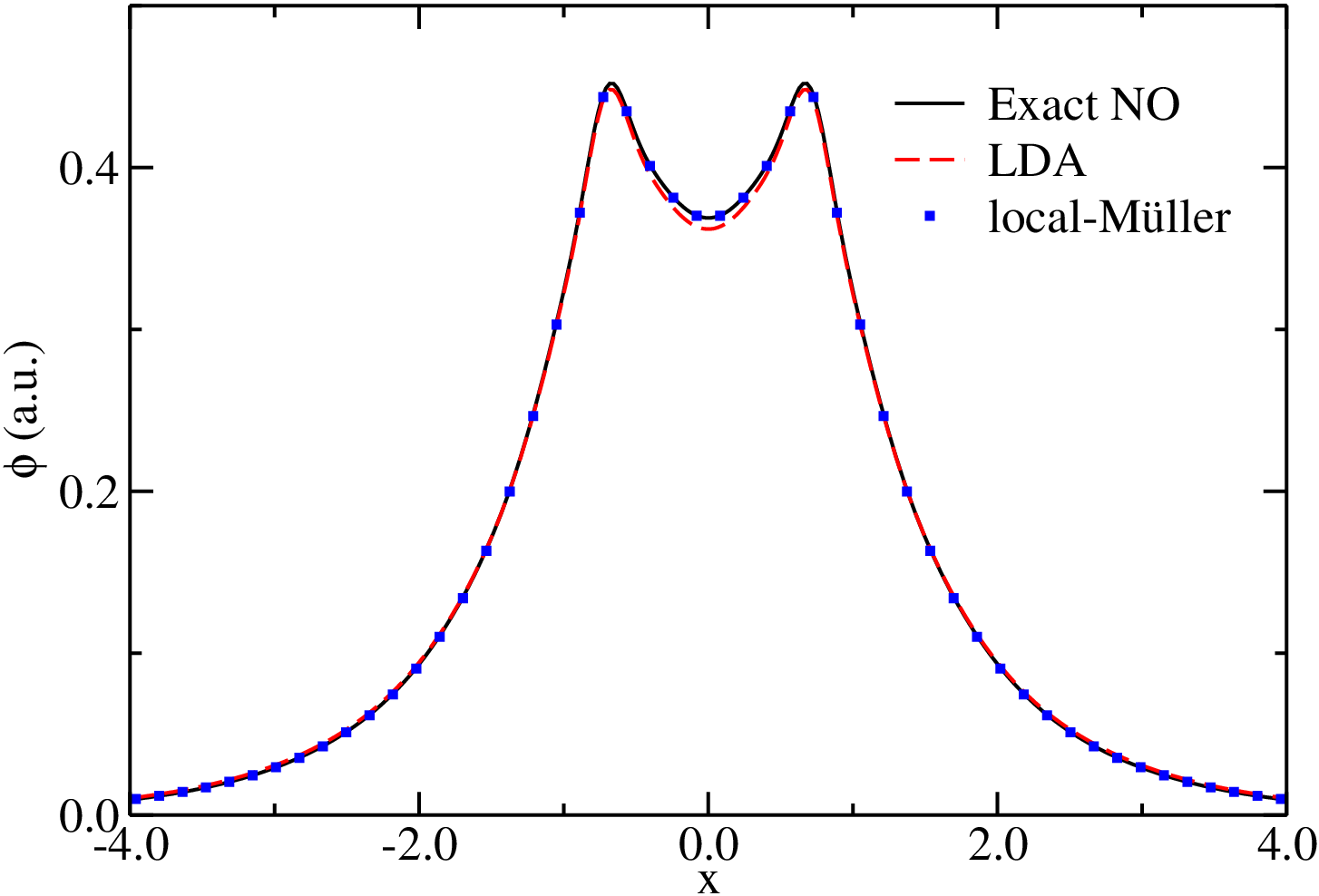} & 
\includegraphics[width=0.45\textwidth, clip]{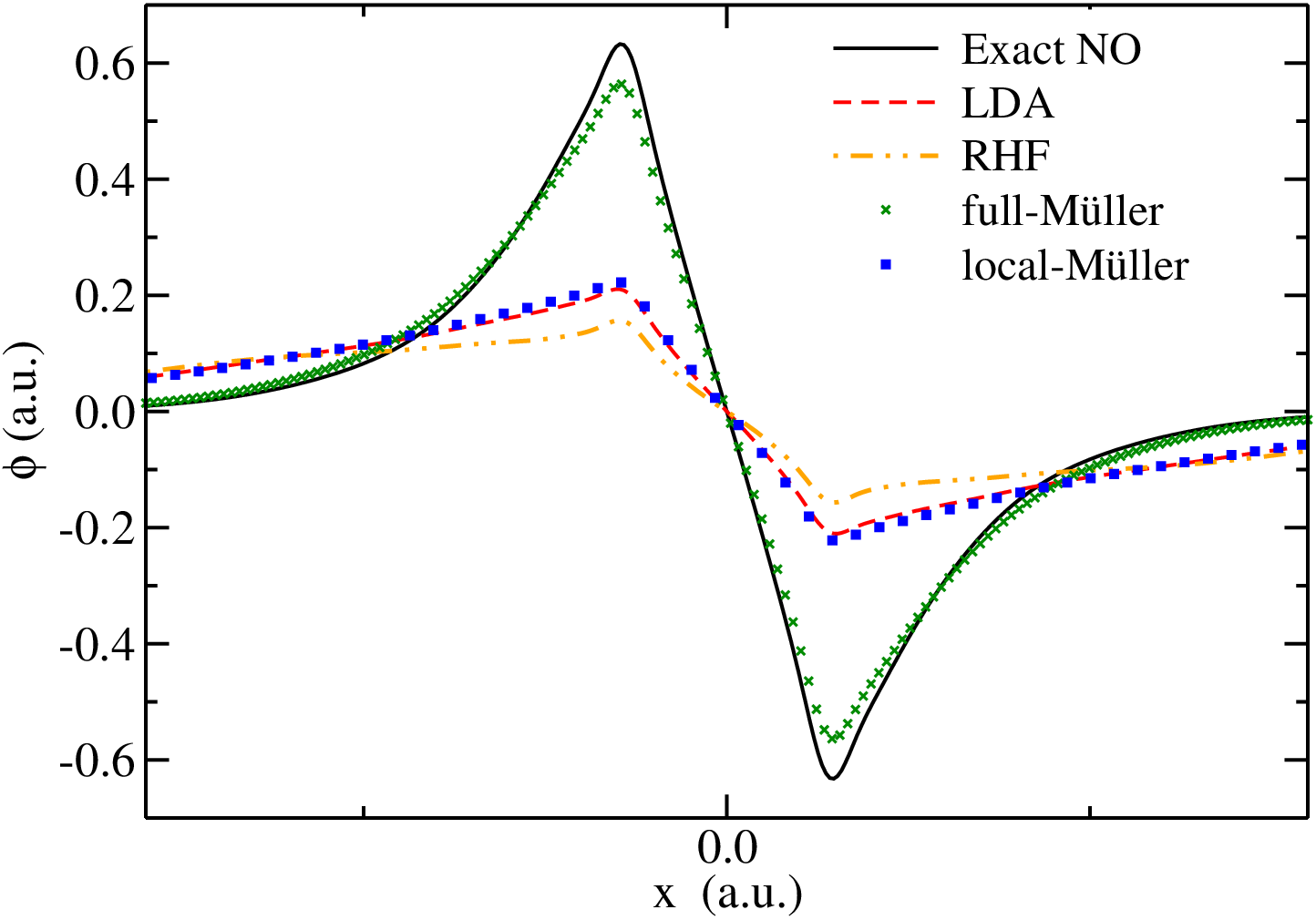} \\
\end{tabular}
\caption{\label{fig:H2_eq} H$_2$ 1$\sigma$ (left) and H$_2$ 1$\sigma^{*}$ (right) orbitals at equilibrium plotted along the nuclear axis. For the 1$\sigma$, all full- and local-RDMFT functionals employed yield the same orbital in the scale of the plot, therefore, only the local-M\"uller orbital is shown which also almost coincides with the exact NO and that from RHF. For the 1$\sigma^{*}$, all local-RDMFT functionals employed yield almost the same orbital, while the full-RDMFT orbitals look slightly different from each other but are all close to the exact natural orbital. Only the full- and local-M\"uller orbital is shown}.
\end{figure*}

The theory of local-RDMFT was presented recently in \cite{localrdmft}, here we give a brief summary of the method and state the equations which are solved numerically.
Commonly, RDMFT functionals depend separately on the occupation numbers $n_j$ and the NOs, $\varphi_j(\v r)$. The total energy is then minimized in a two-step process: first one minimizes with respect to the occupation numbers for a fixed set of NOs and then keeps those occupation numbers fixed while minimizing with respect to the NOs. These two steps are then iterated until convergence is achieved. While the minimization wrt.\ the occupation numbers is numerically inexpensive, the optimization of the NOs is quite costly. The local-RDMFT approach reduces the cost for this minimization significantly while still providing fractional occupation numbers through the usual optimization under Coleman's $N$-representability conditions.

The central idea in local-RDMFT~\cite{localrdmft} is that the search for the set of LROs is 
restricted to the domain of orbital sets that satisfy single-particle equations with a local potential
\be
\left[-\frac{\nabla^2}{2} + V_{\rm ext}(\v r) + V_{\rm rep}(\v r)\right]\phi_j(\v r)=\epsilon_j\phi_j(\v r).
\ee 
The search for the electron-electron repulsive part $V_{\rm rep}({\br})$ of the effective local potential 
(the analogue of the Hartree-exchange and correlation potential in the KS equations) is 
replaced by a search for the effective repulsive density (ERD) 
$\rho_{\rm rep}({\br})$ whose electrostatic potential is $V_{\rm rep}({\br})$, i.e. 
\begin{equation}
\nabla^2 V_{\rm rep}({\br}) = - 4 \pi \rho_{\rm rep}({\br})\,. 
\end{equation}
Minimizing with respect to $\rho_{\rm rep}({\br})$ allows for the implementation of two additional constraints \cite{GL2012}:
\begin{eqnarray}
\int d{\br} \: \rho_{\rm rep}({\br}) &=& N-1, \label{eq:asympt}\\
\rho_{\rm rep}({\br}) &\ge& 0\,. \label{eq:pos}
\end{eqnarray}
The two constraints are sufficient to correct the asymptotic behavior of the effective local potential, thus healing to a large extend the self interaction problem expected to be present in the potential
for many approximate functionals in local-RDMFT.

The optimal ERD and the effective local potential can be obtained, similarly to the OEP method, by solving the integral equation~\cite{localrdmft}:
\begin{equation}\label{eq:rho}
\int d^3 r' \, {\tilde \chi} (\v r , \v r') \, \rho_{\rm rep} (\v r') = {\tilde b}( \v r), 
\end{equation}
with
\begin{equation}
\label{eq:oep2}
{\tilde \chi} ( \v r , \v r')\! \doteq\!\! \iint \!\! d^3 x \, d^3 y \, \frac{ \chi ( \v x , \v y) }{ | \v x - \v r| | \v y - \v r'| }\,, 
\end{equation}
\begin{equation}
{\tilde b} ( \v r ) \! \doteq \int \!\!   d^3 x \, \frac{ b ( \v x ) }{ | \v x - \v r|  }\, .
\end{equation}
The response function $\chi (\v r, \v r')$ and $b (\v r)$ are  given by
\begin{equation}
\label{eq:A}
   \chi (\v r , \v r') = {\sum_{j,k,j \ne k}}  \phi_j^*(\v r)\,\phi_k(\v r)\,\phi_k^*(\v r')\,\phi_j(\v r')\frac{n_j-n_k}{\epsilon_j -\epsilon_k}\,, 
\end{equation}
\begin{equation}
   b(\v r) = {\sum_{j,k,j \ne k}} \langle \phi_j | \frac{F_{\rm Hxc}^{(j)}-F_{\rm Hxc}^{(k)}}{\epsilon_j - \epsilon_k} 
| \phi_k \rangle\,  \phi_k^*(\v r)\, \phi_j(\v r)\,,
\label{eq:B}
\end{equation}
with $F_{\rm Hxc}^{(j)}$ defined by 
\begin{equation}
\label{eq:Liapis}
\frac{\delta E_{\rm Hxc}}{\delta \phi_j^*(\v r)} \doteq
\int d^3r' \, F_{\rm Hxc}^{(j)}(\v r, \v r') \, \phi_j(\v r')\,. 
\end{equation}
$E_{\rm Hxc}$ is the approximation for the electron-electron interaction energy,  $\phi_j$ are the LROs and $n_j$, and $\epsilon_j$ their corresponding occupation numbers and orbital energies (eigenvalues of the effective Hamiltonian).
The two constraints are incorporated with a Lagrange multiplier (\ref{eq:asympt}) and a penalty term (\ref{eq:pos}) that introduces an energy cost for every point $\br$ where $ \rho_{\rm rep}(\br)$ becomes negative.

Within local-RDMFT, the numerical cost for one minimization wrt.\ the NOs reduces to the cost for a DFT calculation using an orbital functional. Apart from reducing the numerical cost, the local-RDMFT framework provides an energy eigenvalue spectrum connected to the LROs. We found that the energy eigenvalues of the strongly occupied orbitals reproduce the ionization potentials of small and large molecules \cite{localrdmft, localrdmftappl} accurately. We now turn our attention to the question of how the LROs compare with different sets of orbitals like the almost exact NOs taken from MCSCF calculations, those from the full RDMFT minimization, and the KS orbitals of LDA.

\section{Results and discussion}\label{sec:results}

In this section, we perform full- and local-RDMFT calculations for the 
He atom, and the H$_2$, molecule at 3 different intermolecular distances, the equilibrium geometry with $R=1.4\: a.u.$,
an intermediate distance with $R=3\:a.u.$, and at the dissociation limit, $R=8\:a.u.$, in order to examine the impact of the 
local potential constraint on the optimized orbitals. For full- and local-RDMFT we employ
a few representative approximations, namely the M\"uller \cite{M1984}, BBC3 \cite{GPB2005}, 
Power \cite{sharma08,power_finite}, ML \cite{pade}, and L\"owdin Schull (LSH) \cite{LSH1956} functionals. The LSH functional is the exact one for the 2-electron singlet case in the full minimization. Apparently, in local-RDMFT, with the assumption of a local potential, LSH is no longer exact and not necessarily better than other approximations. 
 For comparison, we also calculated the NOs obtained within the complete active 
space MCSCF method. As these are very accurate approximations to the exact NOs, we will refer to them as ``exact'' NOs in the following. Full minimization orbitals from the LSH functional obviously 
coincide with those from the MSCF method, since we discuss only two electron singlet systems. Finally, we compare the LROs with restricted HF (RHF) and LDA orbitals. All calculations were performed with the HIPPO computer code \cite{code}, except for MCSCF for which we used the GAMESS code \cite{GAMESS}. In all calculations, the cc-PVTZ Gaussian basis set was employed. For simplicity, and in order to avoid numerical noise introduced in weakly occupied orbitals, we compare orbitals with an occupation larger than $10^{-4}$.
In order to compare orbitals qualitatively we plot them along the radial direction for the He atom and along the intermolecular axis for the H$_2$ molecule. To keep the figures uncluttered we plot only the orbital from one RDMFT functional (full and local), together with the MCSCF, the LDA and the RHF ones. In addition,
as a more quantitative comparison, we calculate the overlaps of the corresponding orbitals obtained with full RDMFT and local-RDMFT, for all the functionals employed, with the exact
NOs and KS-LDA orbitals.

Before discussing the nature of the LROs, we present more evidence that static correlations can be correctly described by
local-RDMFT if the corresponding 1RDM functional has this property. In Fig.~\ref{fig:nd}, we show the occupation numbers 
of the 1$\sigma$ and 1$\sigma^*$ spin-orbitals of the H$_2$ molecule obtained with four different functionals in local-RDMFT. We see that the dependence of these occupations on the distance is qualitatively correct  going to the correct dissociation value of 1/2. Occupations from Local-LSH and local-BBC3, in particular, are quite accurate compared with the exact ones. The ML functional is not included in this plot as it does not yield a qualitatively correct dissociation curve for H$_2$ in either the local or the full RDMFT version. As H$_2$ at large internuclear separation is a prototype correlated system, it is of particular interest to see if the negative of the orbital energy of the HOMO is a good approximation for the ionization potential (IP). In the Table~\ref{tab:ip}, we show the calculated IPs at two internuclear separations, at equilibrium and at a distance of 3 \AA. Results from full RDMFT using the extended Koopman's theorem (EKT) \cite{MPL1975,SD1975,PC2005,PMLU2012,PMLU2013} are also shown. EKT is proven to yield the IP exactly for the exact functional of the 1RDM. It also provides a new orbital representation in terms of the canonical orbitals which can be associated with the electronic spectrum \cite{PMLU2013,MRILUMP2013}. 
Finally, in Table~\ref{tab:ip}, we also include the exact results within the particular basis set which are obtained using the definition of the IP i.e.\ the total energy difference $E(N-1)-E(N)$, where $E$ is the total energy and $N$ is the number of electrons. As expected, the IP of the EKT method is exact for the LSH functional.
Local-LSH and BBC3 functionals on the other hand are very accurate at the equilibrium distance and rather satisfactory
at large separations as well, as shown in the Table~\ref{tab:ip}.

\begin{table}
\setlength{\tabcolsep}{0.29cm}
\begin{tabular}{cccccc}
\hline\hline
{Distance} & \multicolumn{2}{c}{{BBC3}} & \multicolumn{2}{c}{{LSH}} & {Exact} \\
  {(\AA )}        & {Local} & {EKT} & {Local} & {EKT} & \\ \hline
{0.77} &  {15.98} & {15.94} & {15.98} & {16.20} & {16.20} \\
{3.00}       &  {12.75} & {12.44} &{12.89} & {13.26} & {13.26} \\ 
\hline\hline
\end{tabular}
\caption{\label{tab:ip} IPs (in eV) from the HOMO eigenvalues of local-BBC3 and local-LSH for H$_2$ at equilibrium and at 3~\AA\ 
compared with values from EKT with BBC3 and LSH full-RDMFT functionals. The exact values, which are obtained from the energy difference
of the positive ion and the neutral system, are also included.}
\end{table}
We continue with discussing the nature of the LROs. The helium 1$s$ orbital appears to be identical for all methods considered in this work showing the typical shape of a 1$s$ orbital. This orbital is strongly occupied in local- and full-RDMFT calculations and its occupation varies between $1.975$ and $1.994$. The He 2$s$ orbital, on the other hand, shown in Fig.~\ref{fig:He_2nd}, is strongly dependent on the calculation method. The occupation of this orbital varies between $0.0019$ and $0.011$. Within the local-RDMFT approach, all four different functionals yield an almost identical form on the scale of the plot, hence, only the one obtained with the M\"uller functional is included in Fig.\ \ref{fig:He_2nd}. Also, the orbitals from the M\"uller and Power functionals in the full RDMFT calculation are identical, and again, only the one given by M\"uller is shown.
Interestingly, all LROs from local-RDMFT are almost identical with the corresponding LDA orbital and are also close to the RHF one. Orbitals from full-RDMFT are closer to the exact NO with the orbital from BBC3 being the closest.

\begin{table}[h]
\setlength{\tabcolsep}{0.25cm}
\begin{tabular}{l|cc|cc}\hline\hline
& \multicolumn{2}{c|}{local-RDMFT} & \multicolumn{2}{c}{full-RDMFT}\\
& Exact NO & LDA & Exact NO & LDA\\  \hline
M\"uller & 0.85803 & 0.99964 & 0.99657 & 0.89302\\
BBC3 & 0.85748 & 0.99967 & 0.99956 & 0.86825\\
ML & 0.85878 & 0.99953 & 0.98462 & 0.75565\\
Power & 0.85792 & 0.99963 & 0.99620 & 0.89507\\ 
{LSH}   &{ 0.85801} &{ 0.99965} & {0.99999} & {0.85369} \\ \hline\hline
\end{tabular}
\caption{\label{tab:he_2s} Overlaps of the 
 Helium 2$s$ orbital employing different functionals within 
local- or full-RDMFT with the corresponding exact NO and LDA orbital.}
\end{table}

The overlaps of helium 2$s$ orbital obtained with different methods are shown in Table \ref{tab:he_2s}. 
The overlaps of the full-RDMFT orbitals with the exact NO are larger than 0.98 for all approximations. 
On the contrary, the local-RDMFT orbitals for all functionals are almost identical to the LDA orbital with overlaps larger than 0.999.

In Figs.~\ref{fig:H2_eq}(a) and \ref{fig:H2_eq}(b) we plot the 1$\sigma$ and 1$\sigma^*$ orbitals, for the hydrogen molecule at equilibrium, along the intermolecular axis. The first orbital looks exactly the same for both local- and full-RDMFT and all approximate functionals. It is also very close to the exact NO and RHF orbitals while the LDA gives a slightly different orbital which is, however, still close to the NOs. The overlaps between the RDMFT orbitals, both local and full, with the exact NO never deviate from one by more than $3\cdot 10^{-5}$. The overlaps with the LDA orbital are only marginally smaller with a deviation from one of about $3\cdot 10^{-4}$. The antibonding 1$\sigma^{*}$ orbital again looks exactly the same for all local-RDMFT functionals tested and almost coincides with the corresponding LDA orbital. The RHF orbital is also close to them. The full RDMFT NOs are again close to the exact NO, with the BBC3 being the closest. The Power full-RDMFT orbital lies between the M\"uller and the BBC3 orbitals and is omitted from the plot for clarity. These findings are again confirmed by the overlaps between the different sets of orbitals which are given, for the 1$\sigma^*$ orbital, in Table \ref{tab:h2_2nd}: The overlap between the local-RDMFT orbitals and the LDA orbital being around 0.999 in all cases and the full RDMFT orbitals having an overlap with the exact NO larger than 0.99.

\begin{table}
\setlength{\tabcolsep}{0.25cm}
\begin{tabular}{l|cc|cc}\hline\hline
& \multicolumn{2}{c|}{local-RDMFT} & \multicolumn{2}{c}{full-RDMFT}\\
& Exact NO & LDA & Exact NO & LDA\\  \hline
M\"uller & 0.74144 & 0.99892 & 0.99537 & 0.76873\\
BBC3 & 0.73858 & 0.99911 & 0.99954 & 0.71960\\
ML & 0.73719 & 0.99920 & 0.99767 & 0.75372\\
Power & 0.74059 & 0.99896 & 0.99138 & 0.64364\\ 
{LSH}   &{ 0.73759} &{ 0.99917} &{ 0.99999} &{ 0.71255} \\ \hline\hline
\end{tabular}
\caption{\label{tab:h2_2nd} H$_2$ 1$\sigma^*$ orbital at equilibrium: Overlaps between local- and full- RDMFT orbitals with 
exact NO and LDA orbitals for different RDMFT approximations.}
\end{table}

\begin{table}
\setlength{\tabcolsep}{0.25cm}
\begin{tabular}{l|cc|cc}\hline\hline
& \multicolumn{2}{c|}{local-RDMFT} & \multicolumn{2}{c}{full-RDMFT}\\
& Exact NO & LDA & Exact NO & LDA\\  \hline
M\"uller & 0.80117 & 0.99771& 0.93510 &	0.88263\\
BBC3     & 0.79625 & 0.99850& 0.99952 &	0.79624\\
ML       & 0.79297 & 0.99983& 0.95454 &	0.89306\\
Power    & 0.79368 & 0.99975& 0.98611 &	0.75053\\
{LSH} & {0.78874} & {0.99991}& {0.99999} &{0.78008}\\\hline\hline
\end{tabular}
\caption{\label{tab:h2_inter_3rd} H$_2$ 2$\sigma$ orbital at intermediate distance 3 a.u.: Overlaps between local- and full-RDMFT orbitals with exact NO and LDA orbitals for different RDMFT approximations.}
\end{table}

\begin{figure}
\includegraphics[width=0.4\textwidth, clip]{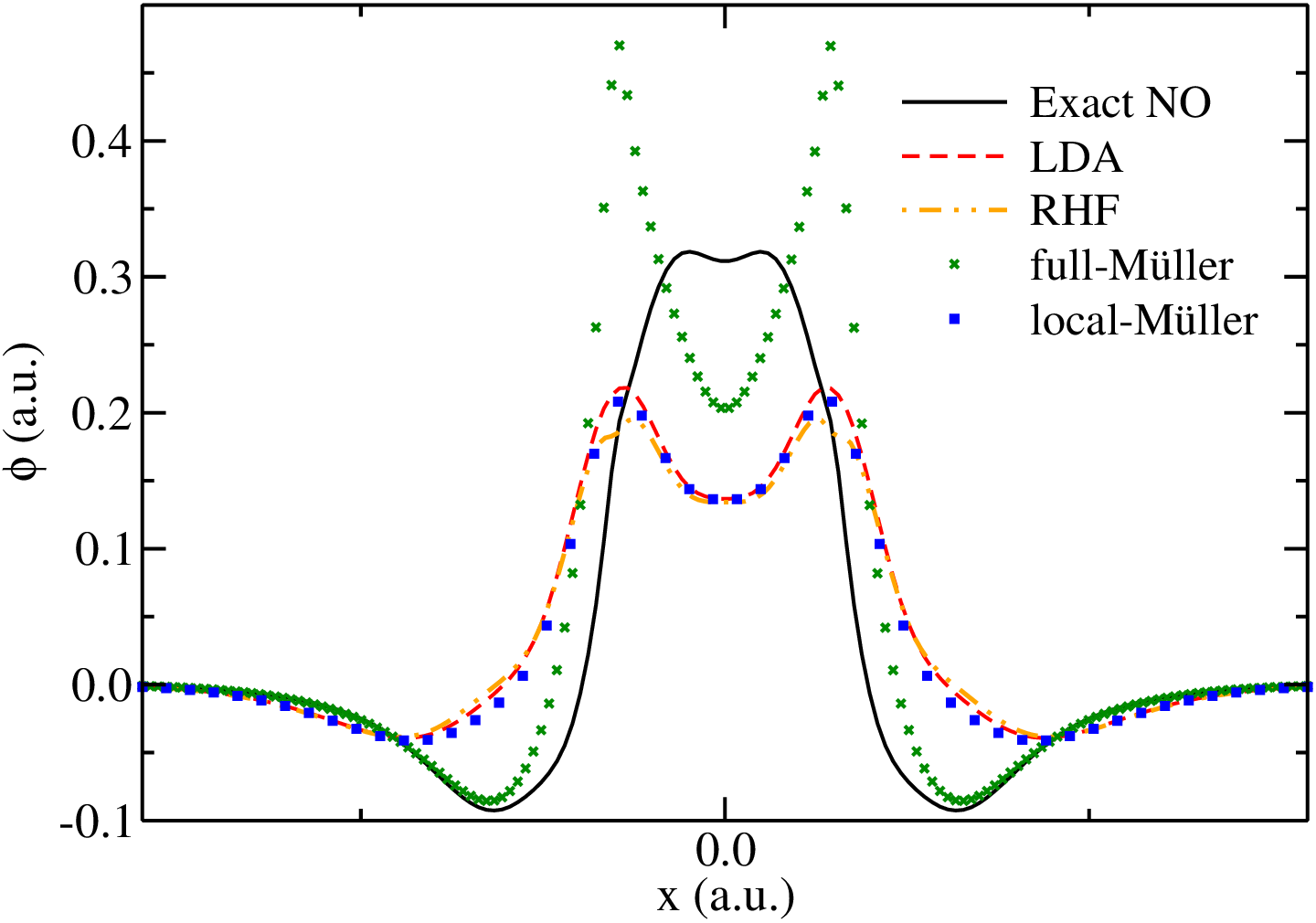}
\caption{\label{fig:H2_3_3rd}  
 H$_2$ 2$\sigma$ at internuclear distance 3 a.u.: All local-RDMFT functionals look almost the same. The full-RDMFT orbitals differ slightly from each other but are all closer to the exact natural orbital than to the LDA and RHF orbitals. Only the local and full-M\"uller orbitals are shown. LDA and RHF orbitals almost coincide}.
\end{figure}
At an internuclear distance of $3\: a.u.$ one finds practically the same orbital for the 1$\sigma$ and 1$\sigma^*$ orbitals for all methods considered in this work. Therefore, they are not discussed in more detail. 
In Fig.\ \ref{fig:H2_3_3rd} we plot instead the 2$\sigma$ orbital for this system, the only orbital, with a significant occupation, which differs depending on the employed method. As in all other cases considered, local-RDMFT gives exactly the same orbital for all functionals. The corresponding LDA and RHF orbitals look similar and are close in shape to the local-RDMFT orbitals. On the other hand, the orbitals obtained by full-RDMFT differ significantly. While the BBC3 full RDMFT orbital is very similar to the exact NO, the Power, the ML and the M\"uller orbitals differ from it especially close to the nuclei. Overall, however, the full RDMFT orbitals are significantly closer to the exact NO than the local-RDMFT ones. This is also reflected in the overlaps given in Table \ref{tab:h2_inter_3rd} which show a large overlap between the full RDMFT orbitals and the exact NO while the local-RDMFT orbitals are rather closer to the LDA orbital than to the exact one.

\begin{figure}
\begin{tabular}{c}
\includegraphics[width=0.4\textwidth, clip]{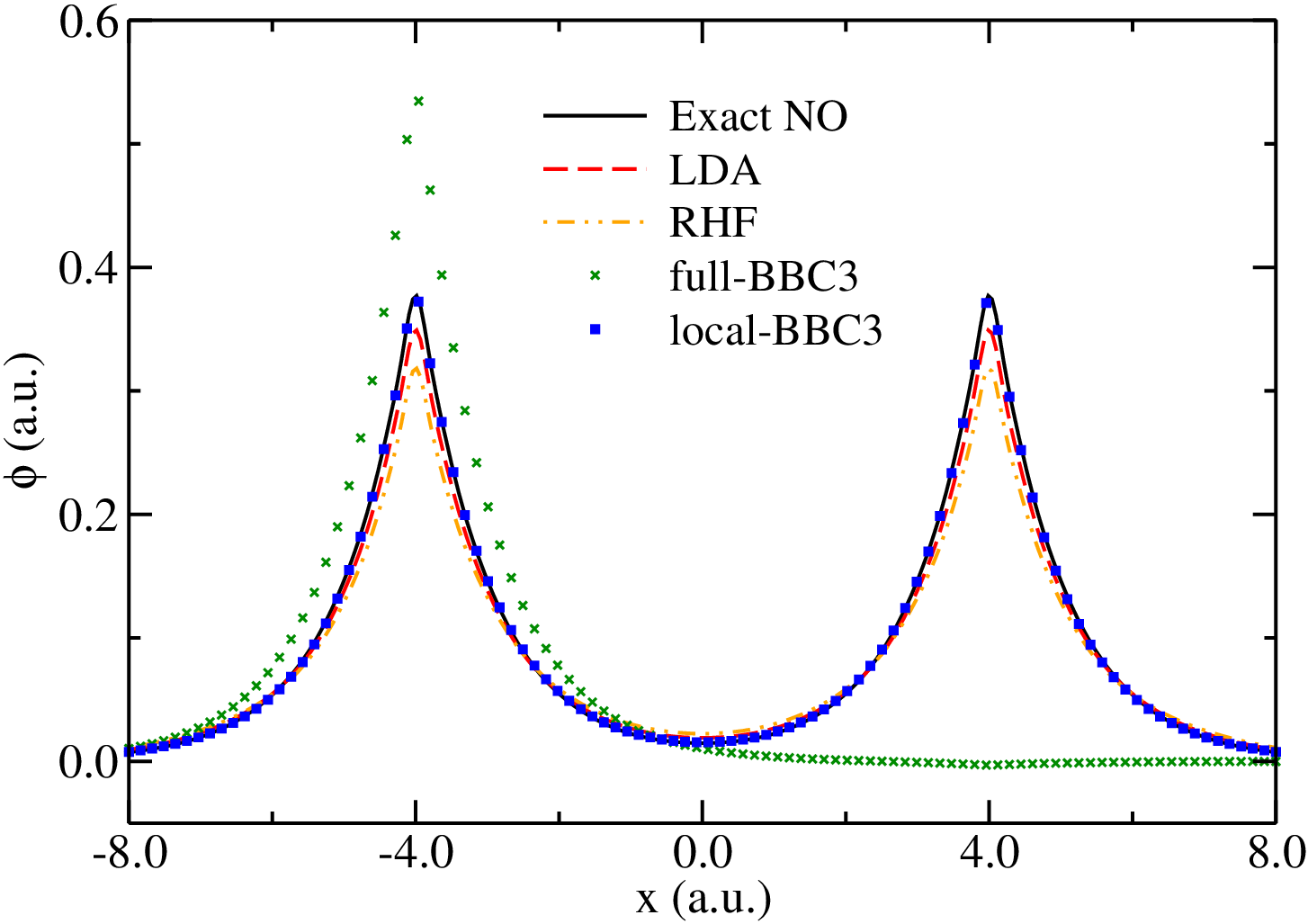} \\
\includegraphics[width=0.4\textwidth, clip]{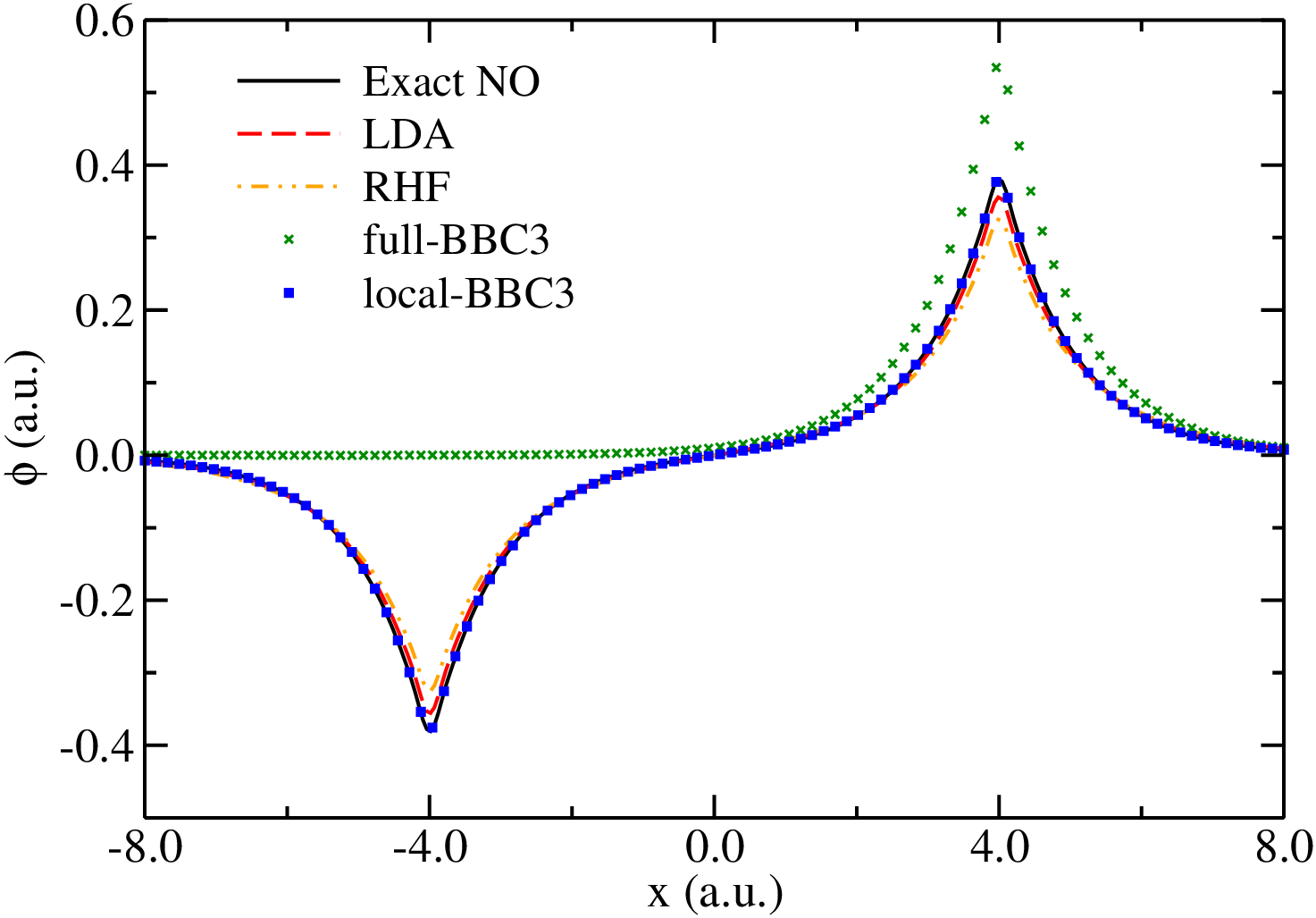} \\
\end{tabular}
\caption{\label{fig:H2_dis}  H$_2$ 1$\sigma$ (top) and H$_2$ 1$\sigma*$ (bottom) orbitals at 
dissociation. Essentially all functionals for full and local-RDMFT
yield a symmetric and antisymmetric superposition of two hydrogenic orbitals except full-BBC3
that gives two localized orbitals instead as discussed in the text. }
\end{figure}

In Figs.\ \ref{fig:H2_dis} top and bottom panels, we show the first two orbitals for the hydrogen molecule at the dissociation limit. Since most orbitals look almost identical on the scale of the plot we only included some examples to show the general behavior. At this limit the two orbitals plotted are the symmetric and antisymmetric combination of hydrogen atom 1$s$ orbitals for all approximations except full-BBC3. The 
reason for this difference is because this approximation is not invariant under a 
delocalization unitary transformation in a subspace of localized NOs with degenerate 
occupation numbers. As it was shown in Ref.~\onlinecite{LGH2010}, BBC3 favors energetically
orbital localization and, in that way, it remains size consistent. 
Local-BBC3, since it is based on the same energy functional, should also favor orbital localization. The reason it 
does not is that the single particle Hamiltonian with the local potential is invariant in the subspaces of degeneracies (an occupation number degeneracy leads to a single particle energy degeneracy). Thus, for the effective Hamiltonian it is equivalent to get localized or delocalized pairs of orbitals and the choice is mostly a technical matter of the diagonalization routine. In our case, the result with local-BBC3 resembles that of the other approximations, i.e.\ a  pair of symmetric and antisymmetric delocalized orbitals. This small discrepancy between the full and the local version of a given functional can be resolved by a subsequent minimization of the original functional in the subspace of degeneracy. Generally, we should keep in mind that the local Hamiltonian is invariant under orbital rotations while the energy functional might not be. In such cases, a small scale additional minimization of the original functional in the degenerate subspace is required.

\begin{table}
\setlength{\tabcolsep}{0.7cm}
\begin{tabular}{lcc}\hline\hline
Functional & $\Delta_{\rm LDA}$ & $\Delta_{\rm cLDA}$ \\ \hline
M\"uller & -6.32 & -1.27 \\
BBC3     & -6.24 & -1.17 \\
ML       & -4.96 & -0.48 \\
Power    & -5.54 & -0.50 \\ \hline\hline
\end{tabular}
\caption{\label{tab:eigs} The average differences (in eV) of the local-RDMFT eigenvalues, for the orbitals 
considered in the present work, from those of the LDA ($\Delta_{\rm LDA}$)  and cLDA 
($\Delta_{\rm cLDA}$).
}
\end{table}

Since for all systems studied here, the local-RDMFT orbitals are close to the LDA ones, one might wonder if the same is true for the corresponding energy eigenvalues. We found this not to be the case and the average differences from  the
LDA eigenvalues for different functionals are of the order of 5-6 eV, for the orbitals considered in this work, as one can see 
in Table~\ref{tab:eigs}.  The relative differences for all functionals
lie in the range of $32\%$ to $463\%$. This is largely expected since local-RDMFT energy eigenvalues
have been proven to give accurate approximations to the experimental IPs~\cite{localrdmft,localrdmftappl} contrary to the LDA ones.
The big difference from the LDA eigenvalues can be partly attributed to the constraints (\ref{eq:asympt}) and (\ref{eq:pos}). 
If one compares the local-RDMFT eigenvalues with those from LDA calculations where the same constraints are imposed \cite{GL2012}
(referred to as constrained LDA (cLDA)) one finds an improved agreement in all cases, as shown in Table~\ref{tab:eigs}. 
The average energy differences in this case are in the range of 0.5-1.3~eV for the different local-RDMFT approximations while
the relative differences for all functionals are in the range of $0.2$ - $64\%$.

The close resemblance of local-RDMFT orbitals to LDA ones suggests that,
 as far as static correlations are concerned, one could completely avoid the orbital optimization. Instead one could perform a single minimization of occupation numbers using KS orbitals, for example from a LDA calculation, and still obtain a good description of static correlation as local-RDMFT was found to provide. In order to obtain non-idempotent solutions for the occupation numbers, it is essential to use 1-RDM approximations for the energy expressions and not density ones \cite{localrdmft}. The resulting method could provide an adequate description of strong static correlation at a computational cost which is only slightly larger than a density functional calculation since the only additional calculation is a one-shot optimization of the occupation numbers which is rather cheap. 
In addition, a single evaluation of the effective potential of local-RDMFT with the modified occupation numbers would offer an improved quasi-particle energy spectrum compared to the initial LDA calculation.

\section{Conclusion}\label{sec:conclusion}

In conclusion, this work focuses on the nature of the optimal orbitals, LROs, of local-RDMFT. This approach was introduced recently as a bridge between RDMFT and DFT. It is based on the idea
of minimizing functionals of the 1-RDM under the additional condition that the  
LROs satisfy a single-particle Hamiltonian with a multiplicative KS-like potential. Apart from
computational efficiency, the motivation for the development of local-RDMFT is that there
exists a single-electron spectrum associated with LROs which was proven quite useful in describing
spectral properties like ionization energies of molecular systems. 

According to the findings of this work, LROs from local-RDMFT are much closer to the Kohn-Sham 
orbitals from LDA than the exact NOs. Orbitals from full-RDMFT on the other hand 
resemble the exact NOs much closer than LROs of local-RDMFT. This fact is a demonstration of the
hybrid nature of local-RDMFT: the combination of non-idempotency, i.e.\ fractional occupancies,
with KS-like orbitals. Interestingly, the nice features of local-RDMFT, like the correct description of molecular dissociation and a single electron spectrum, are combined with KS-like orbitals due to the non-idempotency which 
is introduced through the adoption of functionals of the 1-RDM.

The present work can be seen as the initiative for the quest of approximations that combine 
density functionals leading to KS-like orbitals with approaches based on the 1-RDM that lead to fractional occupancies for these orbitals. Such schemes would combine the simplicity of DFT methods 
with advanced features of RDMFT, like the description of effects that require to go 
beyond the single determinant approximation. The similarity between the local-RDMFT orbitals and KS-LDA orbitals suggests that combining a density functional calculation with a single optimization of the occupation numbers with a 1-RDM functional could provide an inexpensive way for the description of strongly correlated systems in a hybrid
 density/density matrix framework in the future. 

\begin{acknowledgments}
IT and NH acknowledge support from a Emmy-Noether grant from Deutsche Forschungsgemeinschaft. NNL acknowledges support from the Greek Ministry of Education (E$\Sigma\Pi$A program), GSRT action $\rm KPH\Pi I\Sigma$, project ``New multifunctional Nanostructured Materials and Devices - POLYNANO'' (No. 447963). 
AR acknowledges financial support from the European Research Council Advanced Grant DYNamo (ERC-2010- AdG-267374), Spanish Grant (FIS2013-46159-C3-1-P), Grupos Consolidados UPV/EHU del Gobierno Vasco (IT578-13) and European Community FP7 project CRONOS (Grant number 280879-2) and COST Actions CM1204 (XLIC) and MP1306 (EUSpec).
\end{acknowledgments}

\bibliographystyle{apsrev}

\end{document}